\documentclass[lettersize,journal]{IEEEtran}
\usepackage{amsmath,amssymb,amsfonts}
\usepackage{algorithmic}
\usepackage{array}
\usepackage{textcomp}
\usepackage{stfloats}
\usepackage{url}
\usepackage{verbatim}
\usepackage{graphicx}

\usepackage{cite}
\usepackage{wrapfig}
\usepackage{multirow}
\usepackage{orcidlink}
\usepackage[font=footnotesize]{subcaption}
\usepackage[font=footnotesize]{caption}

\usepackage{kotex}

\def\BibTeX{{\rm B\kern-.05em{\sc i\kern-.025em b}\kern-.08em
    T\kern-.1667em\lower.7ex\hbox{E}\kern-.125emX}}
\usepackage{balance}

\begin{document}
\title{Frequency Estimation Based on SNR-adaptive Frequency Estimator Under Wide SNR Range}
\author{
    Hee-Yang~Jung, Dong-Hee~Paek, Woo-Jin~Jung, and~Seung-Hyun~Kong\textsuperscript{*}
    \thanks{Hee-Yang Jung, Woo-Jin Jung and Seung-Hyun Kong are with the Graduate School of CCS Mobility, Korea Advanced Institute of Science and Technology, Daejeon 34051, South Korea (e-mail: skong@kaist.ac.kr; heeyang@kaist.ac.kr; woo-jin.jung@kaist.ac.kr).}
    \thanks{Dong-Hee~Paek is with the Department of Future Mobility, Korea University, Sejong 30019, Republic of Korea (e-mail: dhpaek@korea.ac.kr).
    }%
    \thanks{\textsuperscript{*}: Corresponding author.}
}

\maketitle

\begin{abstract}

Frequency estimation is the problem of estimating individual tone frequencies from noisy multi-tone sinusoidal signals. Existing frequency estimation methods have difficulty accurately estimating both the number of tone frequencies and the individual tone frequencies in low signal-to-noise ratio (SNR) environments, because weak tone frequency components are buried in noise. In addition, existing methods generally exhibit a trade-off between robustness at low SNR and frequency estimation precision at high SNR, making it difficult to achieve consistently superior frequency estimation performance over a wide SNR range. To overcome these limitations, this paper proposes an SNR-adaptive frequency estimator (SAFE). SAFE consists of a time-frequency image neural network (TFINet), which enhances weak tone frequency components at low SNR, and an SNR-based frequency selector (SFS), which selects an appropriate frequency estimator according to the SNR of the estimated tone frequencies. TFINet enhances tone frequency components even in the low-SNR range, while SFS estimates the SNR of each tone frequency and selects either a robust frequency estimator or a super-resolution frequency estimator according to the estimated SNR. This enables SAFE to achieve robustness at low SNR while preserving high precision at high SNR. Simulation results show that SAFE achieves an False Negative Rate (FNR) of 13.00\% over the SNR range from -10 dB to 0 dB, corresponding to an 13.04\% improvement over the state-of-the-art method. In addition, SAFE reduces the Nearest Neighbor-Root Mean Squared Error (NN-RMSE) by 56.67\% compared with the state-of-the-art method, demonstrating that SAFE performs more accurate frequency estimation. Furthermore, experiments using real-world data demonstrate that SAFE provides robust frequency estimation performance even in practical environments with clutter.

\end{abstract}


\section{Introduction}

Estimating individual tone frequencies from noisy multitone sinusoidal signals has long been a fundamental problem in signal processing. Frequency estimation is widely used in applications such as radar imaging \cite{ref1}, communications \cite{ref2}, Global Navigation Satellite System (GNSS) signal acquisition \cite{ref3,ref4,ref5}, target identification \cite{ref6,ref7,ref8}, and underwater acoustics \cite{ref9}. In particular, accurately estimating both the number K of tone frequencies and the individual frequencies from a limited number of samples in the presence of noise is a key factor that determines the performance of practical signal-processing systems.

Existing frequency-estimation methods can be broadly categorized as nonparametric, parametric, and deep learning (DL)-based methods. Nonparametric methods apply a Fast Fourier Transform (FFT) to the input signal to compute the power spectral density (PSD), which is used as the frequency representation (FR) \cite{ref10,ref11,ref12,ref13}. These methods are structurally simple and easy to implement. However, because their frequency resolution depends on the number of input samples, a large number of samples is required for accurate frequency estimation. Consequently, closely spaced tone frequencies are difficult to resolve when the number of available samples is limited.
Parametric methods exploit the signal subspace to alleviate this resolution limitation. In particular, they estimate super-resolution frequencies by using the orthogonality between the signal and noise subspaces \cite{ref14,ref15,ref16}. As a result, parametric methods can achieve more accurate frequency estimation than nonparametric methods with fewer samples. However, obtaining the subspaces requires computationally expensive matrix operations, such as eigenvalue decomposition (EVD) or singular value decomposition (SVD), and the computational burden increases rapidly with the number of samples.

More recently, DL-based methods have been applied to frequency estimation. As the representation capability of DL has been demonstrated across various fields and advances in GPU computing have accelerated both training and inference, neural-network-based frequency-estimation methods have been actively investigated \cite{ref8,ref17,ref18,ref19,ref20}. These methods train a neural network to directly generate a super-resolution FR from a noisy multitone sinusoidal signal. Thus, efficient frequency estimation can be performed without the complex matrix-decomposition operations required by parametric methods.

However, existing methods exhibit degraded frequency-estimation performance in practical low signal-to-noise ratio (SNR) environments. In highly noisy environments, such as underwater acoustics \cite{ref21,ref22} and biomedical signal acquisition \cite{ref23}, weak tone-frequency components can be buried in noise, making it difficult to accurately estimate both the number K of tones and their individual frequencies. Moreover, conventional frequency-estimation methods generally exhibit a tradeoff between robustness at low SNR and precision at high SNR. Methods that are robust to noise at low SNR often lack super-resolution precision at high SNR, whereas highly precise methods at high SNR are vulnerable to noise at low SNR. Jia et al.~\cite{ref24} noted that a method whose noise robustness is valid only over a limited SNR range from -10 to 10 dB can be difficult to apply under more severe conditions, and therefore considered a wider SNR range from -20 to 20 dB to evaluate overall noise robustness. Consequently, existing methods have difficulty providing consistently superior frequency-estimation performance over a wide SNR range.

To address these limitations, this paper proposes an SNR-Adaptive Frequency Estimator (SAFE), designed to provide robust frequency estimation at low SNR while retaining high precision at high SNR. As shown in Fig.~\ref{fig1}, SAFE consists of a robust estimator, the time-frequency image neural network (TFINet); a super-resolution estimator, Prony's Method with Cadzow Denoising (PMCD) \cite{ref14,ref25}; and an SNR-based Frequency Selector (SFS), which selects an appropriate estimator according to the SNR of each tone frequency. TFINet transforms the input signal into a time-frequency image (TFI) and suppresses noise using a neural network designed to enhance tone-frequency components. The SFS estimates the SNR of each tone from the FR generated by TFINet and selects the estimator appropriate for the estimated SNR. In this manner, SAFE provides consistent frequency-estimation performance over a wide SNR range. Simulation results show that, over the SNR range from -10 dB to 0 dB, SAFE achieves an FNR of 13.00\%, corresponding to a 13.04\% reduction relative to the best-performing existing method. SAFE also reduces the nearest-neighbor root-mean-square error (NN-RMSE) by 56.67\% compared with the best-performing existing method, confirming robust estimation in low-SNR conditions. Experiments using real frequency-modulated continuous-wave (FMCW) radar data further demonstrate that SAFE remains more robust than existing methods in practical cluttered environments.
The main contributions of this paper are threefold. First, TFINet is proposed to enhance weak tone-frequency components at low SNR. Second, the SFS is introduced to estimate the SNR of each tone and select an appropriate frequency estimator according to the estimated SNR. Third, extensive simulations and real-world experiments verify that SAFE provides consistently superior frequency-estimation performance over a wide SNR range.

The remainder of this paper is organized as follows. Section II reviews existing frequency-estimation methods and discusses their limitations in low-SNR environments. Section III formulates the noisy multitone sinusoidal signal model considered in this paper. Section IV presents the overall architecture and processing stages of SAFE. Section V evaluates SAFE through comparisons with existing frequency-estimation methods. Finally, Section VI concludes the paper.

\section{Related Work}

This section reviews existing frequency-estimation approaches, including nonparametric, parametric, DL-based, and hybrid methods, and discusses their limitations in estimating both the number K of tones and the individual tone frequencies under low-SNR conditions.

\subsection{Nonparametric Method}

Nonparametric methods use the PSD as an FR. The periodogram \cite{ref13} applies an FFT to the input signal to generate the PSD and estimates the tone frequencies by selecting the K largest peaks. To reduce noise, nonparametric methods such as Bartlett's method \cite{ref10} and Welch's method \cite{ref12} divide the input signal into multiple temporal segments, generate an FR for each segment, and combine the resulting FRs. Although this aggregation suppresses noise, it also degrades frequency resolution \cite{ref26}. Even when the periodogram is used alone, a large number of samples is required to sufficiently mitigate the effect of noise. In particular, frequency-estimation performance degrades at low SNR when the number of input samples is limited.

\subsection{Parametric Method}

Parametric methods decompose the signal and noise subspaces by exploiting their orthogonality and estimate super-resolution tone frequencies using a predefined number of tones. MUltiple SIgnal Classification (MUSIC) \cite{ref15} assumes that the input signal is sinusoidal and employs more steering vectors than the number of input samples to obtain a super-resolution FR compared with nonparametric methods. Estimation of Signal Parameters via Rotational Invariance Technique (ESPRIT) \cite{ref16} similarly separates the signal and noise subspaces, but estimates the tone frequencies without generating an FR and therefore provides higher computational efficiency than MUSIC. Other parametric approaches estimate super-resolution frequencies while suppressing noise; one such method is PMCD \cite{ref14,ref25}. PMCD first reduces noise in the input signal using Cadzow denoising, constructs a Toeplitz matrix, and then separates the signal and noise subspaces through SVD to estimate the tone frequencies. At low SNR, however, noise contaminates the signal subspace and prevents clear separation of the signal and noise subspaces, thereby degrading frequency-estimation performance.

\subsection{Deep Learning-based Method}

Deep learning (DL)-based methods employ neural networks that map an input signal to an FR. Owing to their strong representation-learning capability, these methods have also been studied for two-dimensional frequency estimation \cite{ref17} and synthetic aperture radar (SAR) super-resolution \cite{ref27}. DeepFreq \cite{ref8} was the first to apply DL to super-resolution frequency estimation, while cResFreq \cite{ref18} introduced a complex-valued neural-network architecture tailored to in-phase/quadrature (I/Q) input signals. HResFreqNet \cite{ref20} reduces the influence of noise using an autoencoder architecture, whereas SwinFreq and CVSwinFreq \cite{ref28} employ Transformer-based frequency-estimation models \cite{ref29}. DL-based methods learn to distinguish the nonlinear relationship between tone-frequency components and noise, thereby suppressing noise while improving frequency resolution. However, their performance depends on the quality of the training data; when low-SNR input signals are used for training, dominant noise components can degrade frequency-estimation performance.

\subsection{Hybrid Method}

Hybrid methods combine different estimation techniques to compensate for the limitations of individual methods. Zhang et al.~\cite{ref30} combined coarse estimation, averaging, outlier removal, and refined phase estimation to achieve stable frequency estimation at low SNR. Liao and Chen~\cite{ref31} reduced estimation bias by selecting either the Candan estimator or the phase-corrected Quinn estimator according to the frequency offset. However, both methods target single-tone signals. In contrast, this paper considers multitone signals, estimates the SNR of each tone, and adaptively selects RFE or PC according to the SNR, thereby jointly achieving robustness and precision over a wide SNR range.

\section{Problem formulation}

This section mathematically formulates the frequency-estimation problem considered in this paper and defines the associated notation. A continuous-time signal composed of K tone frequencies can be represented as a linear combination of complex exponentials, as given in Eq.~\eqref{eq1}.
\begin{equation} \label{eq1}
    s(t) = \sum_{k=1}^{K} a_k \exp (j 2 \pi f_k t), t \in \mathbb{R},
\end{equation}
where $f_k \in [0,1]$ denotes the normalized frequency of the $k$ th tone, and $a_k \in \mathbb{C}$ denotes its complex amplitude, which contains both magnitude and phase. With sampling frequency $f_s$, the continuous-time signal $s(t)$ can be represented as a finite-length discrete-time signal, as given in Eq.~\eqref{eq2} \cite{ref32}.
\begin{equation} \label{eq2}
    s[n] = s(nT_s) = \sum_{k=1}^{K} a_k \exp (j 2\pi \frac{f_k}{f_s}n),
\end{equation}
where $n=0,1,\ldots,N-1$, $N$ denotes the number of samples, and $T_s=1/f_s$. The noisy discrete-time signal $\tilde{s}[n]$ is then expressed as in Eq.~\eqref{eq3}.
\begin{equation} \label{eq3}
    \tilde{s}[n] = s[n] + \omega[n],    
\end{equation}
where $\omega[n]$ denotes complex additive white Gaussian noise with zero mean and variance $\sigma^2$. Thus, for $\omega[n]$, the probability density function $p(\omega[n])$ is given by Eq.~\eqref{eq4}.
\begin{equation} \label{eq4}
    p(\omega[n]) = \frac{1}{\pi \sigma^2} \exp \left( -\frac{|\omega[n]|^2}{\sigma^2} \right).
\end{equation}

This paper addresses the problem of jointly estimating the total number $K$ of tone frequencies and the individual tone frequencies $f_k$ from the one-dimensional noisy signal $\tilde{s}[n]$.

\section{Proposed Method}

\begin{figure}[!t]
\centering
\includegraphics[width=\linewidth]{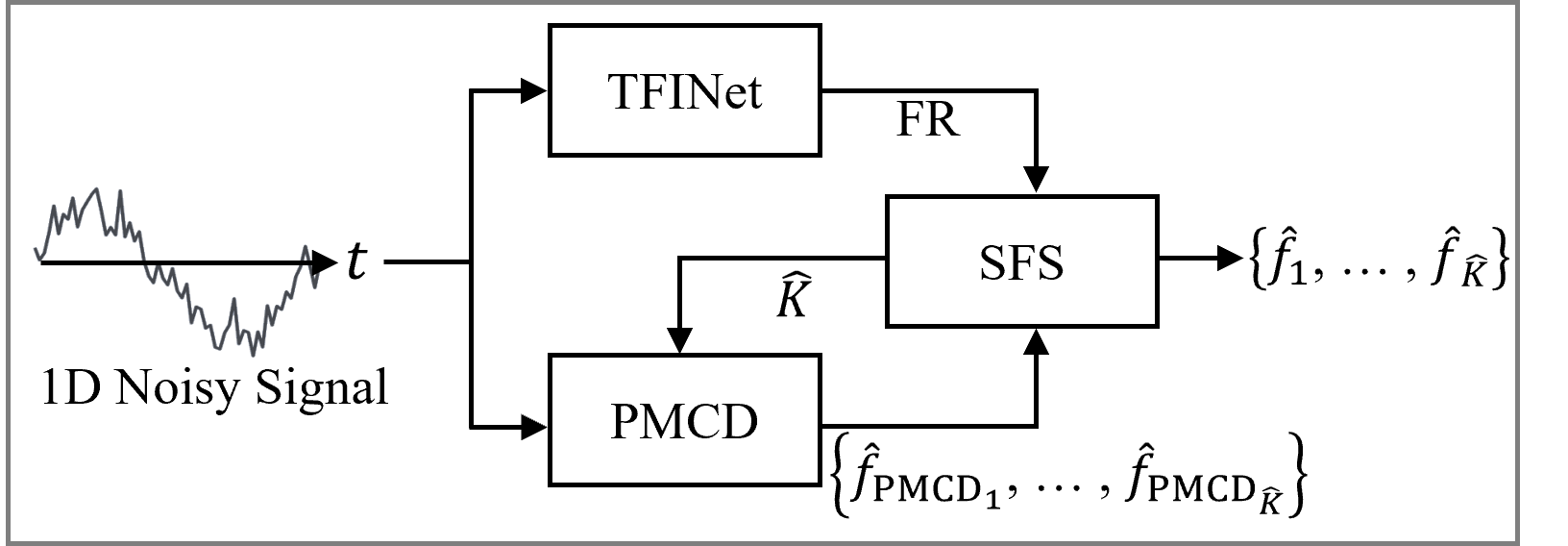}
\caption{Overview of SAFE. SAFE takes a one-dimensional (1D) noisy signal as input and outputs the estimated frequency set $\{\widehat{f}_k\}_{k=1}^{\widehat{K}}$. TFINet generates a noise-suppressed frequency representation (FR), from which SFS detects the tone frequencies. PMCD takes the estimated number of frequencies, $\widehat{K}$, and the 1D noisy signal as inputs and outputs the frequency estimates $\{\widehat{f}_{\mathrm{PMCD},k}\}_{k=1}^{\widehat{K}}$. In SFS, the SNR of each detected frequency is estimated, and either the frequency detected by TFINet or the corresponding estimate $\widehat{f}_{\mathrm{PMCD},k}$ is selected.}
\label{fig1}
\end{figure}

This section describes TFINet and SFS, the two principal components of SAFE. As shown in Fig.~\ref{fig1}, the input signal first passes through TFINet, which enhances weak tone-frequency components. The SFS then estimates the SNR of each tone and selects either the robust estimator TFINet or the super-resolution estimator PMCD according to the estimated SNR.

\subsection{Time-Frequency Image Neural Network (TFINet)}

\begin{figure*}[!t]
\centering
\includegraphics[width=\linewidth]{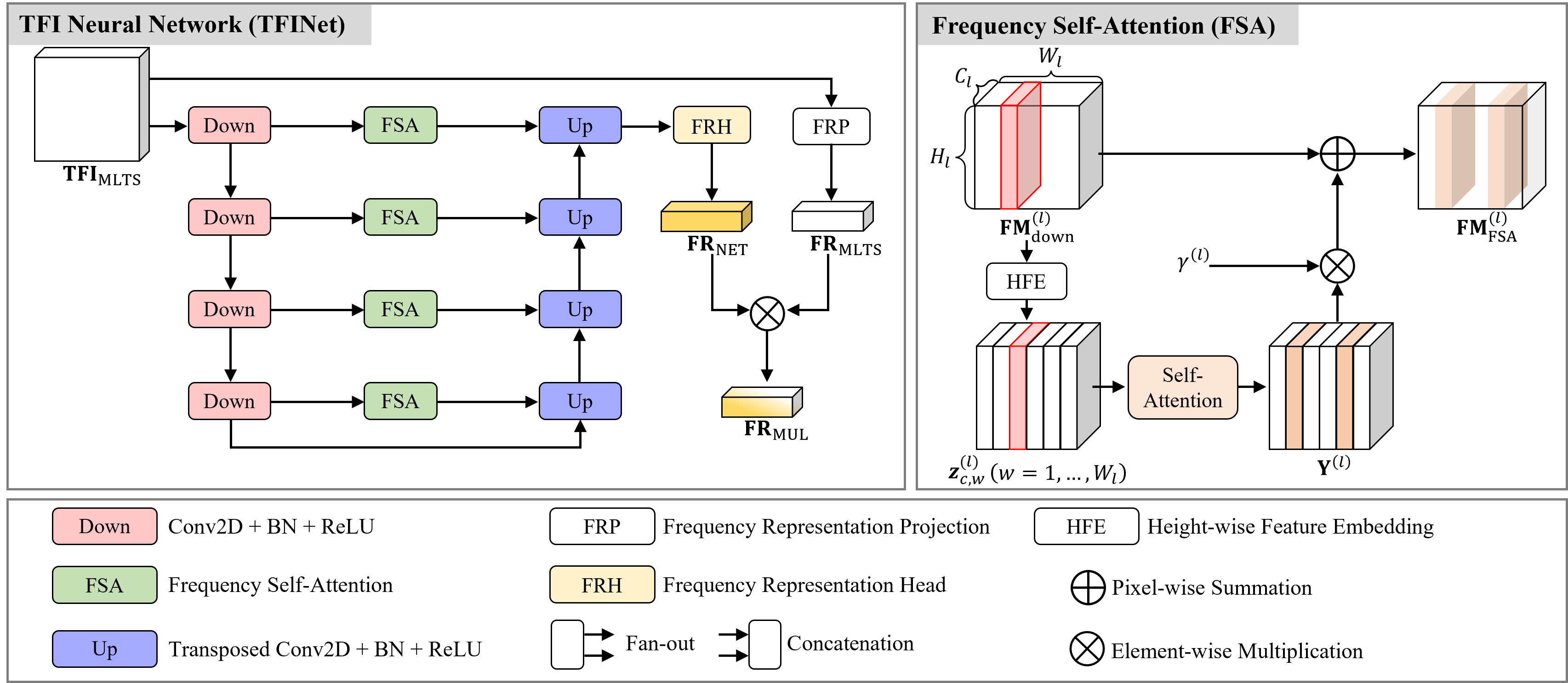}
\caption{Architecture of TFINet. TFINet takes $\mathbf{TFI}_{\mathrm{MLTS}}$ as input and outputs the noise-suppressed frequency representation $\mathbf{FR}_{\mathrm{MUL}}$. In FSA, self-attention is performed along the frequency axis for each feature map $\mathbf{FM}_{\mathrm{down}}^{(l)}$ obtained from the corresponding downsampling layer.}
\label{fig2}
\end{figure*}

TFINet takes a TFI as input and outputs a noise-suppressed FR. As the
input, this study uses
\(\mathbf{TFI}_{MLTS} \in \mathbb{R}^{N \times \left\lfloor N/c_{f} \right\rfloor}\)
proposed by Kong et al.~\cite{ref33}. \(\mathbf{TFI}_{MLTS}\) is generated by
multiplying the input signal \(\widetilde{s}[ n]\) in Eq.~\eqref{eq3} by
multiple linear-frequency-modulated (LFM) signals with different chirp
rates and converting each product into a TFI. In the TFI, tone-frequency
components appear as oblique lines; therefore, an affine transform (AT)
is applied to align them parallel to the time axis. The invalid regions
are then cropped, and the aligned TFIs are summed. Consequently,
tone-frequency components accumulate at the same locations, whereas
noise components are distributed differently across the TFIs and are
therefore relatively suppressed. Here, \(c_{f}\) denotes the cropping
ratio used to remove invalid TFI regions introduced by the AT. Because
some high-amplitude noise components still remain in
\(\mathbf{TFI}_{MLTS}\) and may mask weak tone-frequency components,
TFINet further suppresses the residual noise.

As shown in Fig.~\ref{fig2}, TFINet adopts a U-Net architecture based on a
two-dimensional convolutional neural network (CNN) \cite{ref34} and is trained to detect tone-frequency components that are not
clearly captured in the TFI. TFINet consists of downsampling, frequency
self-attention (FSA), and upsampling stages. The downsampling stage
constructs layers as in Eq.~\eqref{eq5} to extract a high-dimensional feature map
\(\mathbf{FM}_{down}^{(l)} \in \mathbb{R}^{C_{l} \times H_{l} \times W_{l}}\).
The index \(l\) denotes the network-layer index (i.e., the \(l\)th
layer), while \(C_{l}\), \(H_{l}\), and \(W_{l}\) denote the number of
channels, height, and width, respectively, of the feature map output by
layer \(l\).

\begin{equation} \label{eq5}
\mathbf{FM}_{\mathrm{down}}^{(l)} = \mathrm{ReLU}\left( \mathrm{BN}\left( \mathrm{Conv2D}\left( \mathbf{FM}_{\mathrm{down}}^{(l - 1)} \right) \right)\  \right),
\end{equation}

where \(l = 1,2,3,4\) and
\(\mathbf{FM}_{\mathrm{down}}^{(0)} = \mathbf{TFI}_{\mathrm{MLTS}}\). The
operators \(\mathrm{Conv2D}( \cdot )\), \(BN( \cdot )\), and \(\mathrm{ReLU}( \cdot )\) denote two-dimensional convolution, batch normalization, and the
Rectified Linear Unit activation function, respectively. The feature maps \(\mathbf{FM}_{\mathrm{down}}^{(l)}\), which contain information at different scales and structural levels, are retained for use during upsampling.

Each \(\mathbf{FM}_{\mathrm{down}}^{(l)}\) is passed through the FSA module. As
illustrated on the right side of Fig.~\ref{fig2}, FSA is designed to enhance
features corresponding to tone frequencies along the frequency axis. In
a TFI, tone-frequency components form straight structures parallel to
the time axis, whereas noise components do not exhibit consistent
structural patterns. To exploit this difference, FSA represents the
feature formed along the time axis at each frequency bin as a single
embedding vector. The feature map extracted from the \(l\)th
downsampling layer is defined as in Eq.~\eqref{eq6}.

\begin{equation} \label{eq6}
\mathbf{X}^{(l)}= \mathcal{R}\left( \mathbf{FM}_{\mathrm{down}}^{(l)} \right) \in \ \mathbb{R}^{C_{l} \times W_{l} \times H_{l}},\
\end{equation}

where \(\mathcal{R}( \cdot )\) denotes a reshape operator that changes
the dimensions of a feature map, and \(H_{l}\) and \(W_{l}\) correspond
to the time and frequency axes of the TFI, respectively. FSA treats each
frequency bin as a token. Thus, given channel index \(c\), the embedding
vector corresponding to the \(w\)th frequency bin is defined as in Eq.~\eqref{eq7}.

\begin{equation} \label{eq7}
\mathbf{z}_{c,w}^{(l)} = \mathbf{X}_{c,w,:}^{(l)} \in \mathbb{R}^{H_l}, \quad w=1,\ldots,W_l.
\end{equation}

In other words, the feature vector formed along the time axis within a
single frequency bin is used as the embedding vector for that bin.
Stacking the embedding vectors of all frequency bins along the frequency
axis yields the token matrix in Eq.~\eqref{eq8}.

\begin{equation} \label{eq8}
\mathbf{Z}_{c} = \left[\mathbf{z}_{c,1}^{(l)}, \mathbf{z}_{c,2}^{(l)}, \ldots, \mathbf{z}_{c,W_l}^{(l)}\right] \in \mathbb{R}^{W_l \times H_l}.
\end{equation}

As shown in Fig.~\ref{fig2}, FSA uses the same token matrix in Eq.~\eqref{eq8} as the query
\(\mathbf{Q}\), key \(\mathbf{K}\), and value \(\mathbf{V}\). It then
computes the attention scores between frequency bins using scaled
dot-product attention followed by a Softmax operation, as given in Eq.~\eqref{eq9}.

\begin{equation} \label{eq9}
\mathbf{A}_{c}^{(l)} = \operatorname{Softmax}\!\left(\frac{\mathbf{Q}\mathbf{K}^{T}}{\sqrt{H_l}}\right) \in \mathbb{R}^{W_l \times W_l}, \quad \mathbf{Q}=\mathbf{K}=\mathbf{V}=\mathbf{Z}_{c}.
\end{equation}

Through the self-attention operation in Eq.~\eqref{eq9}, high attention is assigned
to tone-frequency features that exhibit a consistent structure along the
time axis, while the influence of irregular noise features is
suppressed. Applying the attention scores from all channels to the
feature map \(\mathbf{X}^{(l)}\) yields Eq.~\eqref{eq10}.

\begin{equation} \label{eq10}
\mathbf{Y}^{(l)} = \mathbf{A}^{(l)}\mathbf{X}^{(l)} \in \mathbb{R}^{C_l \times W_l \times H_l}, \quad \mathbf{A}^{(l)} = \left[\mathbf{A}_{1}^{(l)}; \ldots; \mathbf{A}_{C_l}^{(l)}\right].
\end{equation}

where \([ \cdot \ ;\  \cdot \ ]\) denotes concatenation. To
mitigate the gradient-vanishing problem in deep learning, a skip
connection \cite{ref35} is applied, yielding the final FSA output in
Eq.~\eqref{eq11}.

\begin{equation} \label{eq11}
\mathbf{FM}_{\mathrm{FSA}}^{(l)} = \mathbf{X}^{(l)} + \gamma^{(l)}\mathcal{R}\left( \mathbf{Y}^{(l)} \right),
\end{equation}

where \(\gamma^{(l)}\) is a learnable scalar parameter that controls the
contribution of the attention feature. In this manner, FSA preserves the
characteristics of the original downsampled features while enhancing
features corresponding to tone frequencies.

The upsampling stage progressively restores the spatial resolution
reduced during downsampling and generates a feature map with the same
resolution as the input TFI \(\mathbf{TFI}_{\mathrm{MLTS}}\). Let
\(\mathbf{FM}_{\mathrm{down}}^{(4)} = \mathbf{FM}_{\mathrm{up}}^{(4)}\) at the deepest layer; then, the upsampling operation at each layer is given by Eq.~\eqref{eq12}.

\begin{equation} \label{eq12}
\mathbf{FM}_{\mathrm{up}}^{(l - 1)} = \mathrm{ReLU}\left( \mathrm{BN}\left( \mathrm{TConv2D}\left( \left[ \mathbf{FM}_{\mathrm{up}}^{(l)};\mathbf{FM}_{\mathrm{FSA}}^{(l)} \right] \right) \right) \right),
\end{equation}

where \(\mathrm{TConv2D}( \cdot )\) denotes a two-dimensional transposed
convolution. By concatenating \(\mathbf{FM}_{\mathrm{up}}^{(l)}\) and
\(\mathbf{FM}_{\mathrm{FSA}}^{(l)}\), the upsampling stage exploits feature maps
extracted from both the downsampling stage and FSA.

The frequency-representation head (FRH) then generates an FR from
\(\mathbf{FM}_{\mathrm{up}}^{(0)}\). To suppress noise components, the FRH
projects the TFI along the time axis, as given in Eq.~\eqref{eq13}.

\begin{equation} \label{eq13}
\mathbf{FR}_{\mathrm{proj}}[c,m] = \frac{1}{H_l}\sum_{n=1}^{H_l} \mathbf{FM}_{\mathrm{up}}^{(l)}[c,n,m] \in \mathbb{R}^{H_l \times W_l},
\end{equation}

where \([\  \cdot \ ,\ \  \cdot \ ,\ \  \cdot \ ]\) denotes
indexing along the channel, time, and frequency dimensions,
respectively. In addition, \(c\), \(n\), and \(m\) denote the channel,
time, and frequency indices, respectively. In Eq.~\eqref{eq13},
\(\mathbf{FR}_{proj}\) is passed through a one-dimensional CNN layer and
converted, through a Softmax operation, into the two-channel normalized
FR
\(\mathbf{FR}_{\mathrm{NET}} \in \mathbb{R}^{2 \times \left\lfloor N/c_{f} \right\rfloor}\).

\begin{equation} \label{eq14}
\mathbf{FR}_{\mathrm{NET}} = \mathrm{Softmax}\left( \mathrm{BN}\left( \mathrm{Conv1D}\left( \mathbf{FR}_{\mathrm{proj}} \right) \right) \right),
\end{equation}

Through Eq.~\eqref{eq14}, TFINet formulates the determination of whether each
frequency bin contains a tone as a binary segmentation problem. Specifically, \(\mathbf{FR}_{\mathrm{NET}}[ 0,m]\) and \(\mathbf{FR}_{\mathrm{NET}}[ 1,m]\) denote the probabilities that the \(m\)th frequency bin belongs to the tone-frequency and noise classes, respectively, and the two probabilities sum to one. Accordingly, in Eq.~\eqref{eq15}, the ground-truth FR \(\mathbf{FR}_{\mathrm{GT}}\) is defined as a binary label that is assigned 1 when a bin corresponds to a tone frequency and 0 otherwise.

\begin{equation} \label{eq15}
\mathbf{FR}_{\mathrm{GT}}[m] = \begin{cases}
1, & \text{if the } m\text{th bin is a tone frequency},\\
0, & \text{otherwise}.
\end{cases}
\end{equation}

Because only a small fraction of the large number of frequency bins
corresponds to tone frequencies, a severe class imbalance arises. To
alleviate this issue, this study uses class-balanced binary
cross-entropy \cite{ref36}, normalized by the number of samples
in each class, as the loss function.

Finally,
\(\mathbf{FR}_{\mathrm{MUL}} \in \mathbb{R}^{\left\lfloor N/c_{f} \right\rfloor}\) is obtained by element-wise multiplication of the time-axis projection \(\mathbf{FR}_{\mathrm{MLTS}}\) of \(\mathbf{TFI}_{\mathrm{MLTS}}\) and the tone-enhanced FR \(\mathbf{FR}_{\mathrm{NET}}[ 0,:]\) from
\(\mathbf{FR}_{\mathrm{NET}}\). This operation suppresses the overall noise
components in \(\mathbf{FR}_{\mathrm{MLTS}}\) and attenuates noise-induced peaks
in \(\mathbf{FR}_{\mathrm{NET}}\).

\subsection{Prony's Method with Cadzow Denoising (PMCD)}

This subsection summarizes PMCD, the super-resolution
frequency-estimation method selected by the SFS at high SNR, as
illustrated in Fig.~\ref{fig1}.

Prony's method \cite{ref14} estimates the exponential
components, i.e., the tone frequencies, from a high-order polynomial
whose roots correspond to the finite complex exponential components
\(\left\{ \exp\left( j2\pi f_{k}/f_{s} \right) \right\}_{k = 1}^{K}\)
that constitute the input signal \(\widetilde{s}[ n]\). By
accurately estimating each complex exponential component,
Prony's method enables super-resolution frequency
estimation. However, its estimation performance degrades even in the
presence of a small amount of noise. To address this limitation, Cadzow
denoising \cite{ref25} is applied before Prony's method
to remove noise components from the signal subspace.

At low SNR, however, noise components mask the tone-frequency components
and degrade the frequency-estimation performance of PMCD \cite{ref37}. Therefore, according to the SNR estimated by the SFS,
TFINet is selected at low SNR, whereas PMCD is selected at high SNR.

\subsection{SNR-Based Frequency Selector (SFS)}

\begin{figure}[!t]
\centering
\includegraphics[width=0.8\linewidth]{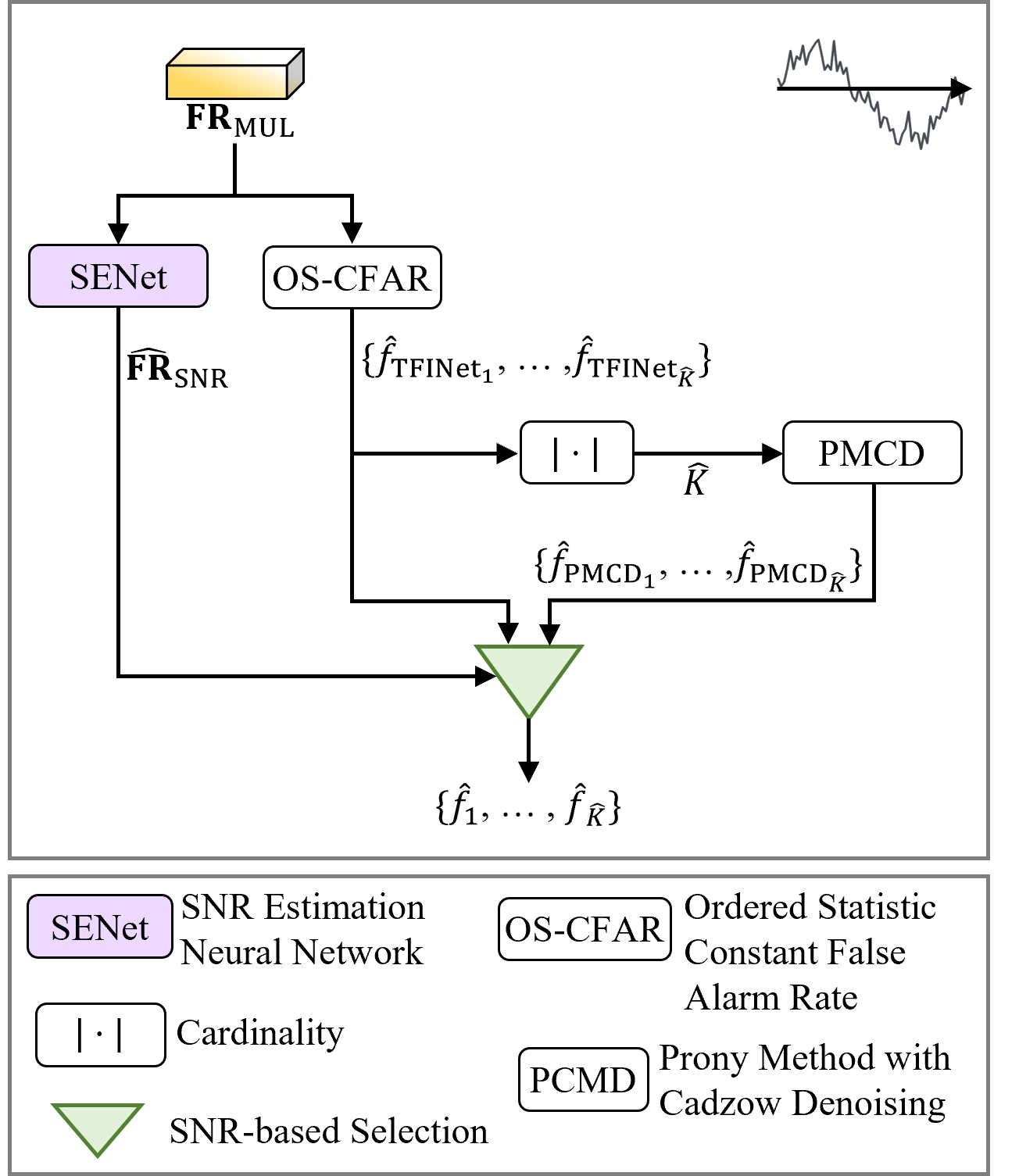}
\caption{Architecture of SFS. SFS estimates $\widehat{\mathbf{FR}}_{SNR}$, an FR representing the SNR information, from $\mathbf{FR}_{MUL}$, and detects the frequency estimates $\{\widehat{f}_{\mathrm{TFINet},k}\}_{k=1}^{\widehat{K}}$ using OS-CFAR. The estimated number of detected frequencies, $\widehat{K}$, is provided to PMCD to obtain the set of super-resolution frequency estimates $\{\widehat{f}_{\mathrm{PMCD},k}\}_{k=1}^{\widehat{K}}$. The SNR of each detected frequency is then estimated from $\widehat{\mathbf{FR}}_{SNR}$, and either $\widehat{f}_{\mathrm{TFINet},k}$ or the corresponding $\widehat{f}_{\mathrm{PMCD},k}$ is selected according to the estimated SNR.}
\label{fig3}
\end{figure}

This subsection describes the SFS, which selects an appropriate
frequency-estimation method over a wide SNR range. As shown in Fig.~\ref{fig1},
the SFS estimates the tone frequencies and their number \(\widehat{K}\)
from the FR generated by TFINet and provides this information to PMCD.
The SFS also estimates the SNR of each tone and selects either the
TFINet output or the PMCD output according to the estimated SNR. The
overall architecture of the SFS is shown in Fig.~\ref{fig3}.

First, the SNR Estimation Neural Network (SENet) produces the SNR
representation \({\widehat{\mathbf{FR}}}_{\mathrm{SNR}} \in \mathbb{R}^{\left\lfloor N/c_{f} \right\rfloor}\) from \(\mathbf{FR}_{\mathrm{MUL}}\), as given in Eq.~\eqref{eq16}.

\begin{equation} \label{eq16}
{\widehat{\mathbf{FR}}}_{\mathrm{SNR}} = \left[ {\widehat{\lambda}}_{0},{\widehat{\lambda}}_{1},\ldots,{\widehat{\lambda}}_{\left\lfloor N/c_{f} \right\rfloor - 1} \right],
\end{equation}

Here, \({\widehat{\lambda}}_{i}\) denotes the estimated SNR at the
\(i\)th frequency bin. SENet employs a Transformer-based model \cite{ref29} to effectively capture tone-frequency peaks. The processing procedure follows Eqs.~\eqref{eq7}--\eqref{eq11}; however, a learnable linear transformation
is applied to \(\mathbf{FR}_{\mathrm{MUL}}\) to expand the limited feature
dimension. Learnable linear transformations are also used when
generating the query, key, and value.

Training SENet requires the ground-truth SNR representation \(\mathbf{FR}_{\mathrm{SNR}}\) corresponding to each tone frequency. To obtain
this representation, let \([ m]\) denote the discrete Fourier transform (DFT) of the noise-free signal \(s[n]\),
as defined in Eq.~\eqref{eq17}.

\begin{equation} \label{eq17}
S[ m] = \sum_{n = 0}^{N - 1}{s[ n]\exp\left( - \frac{j2\pi nm}{N} \right)},
\end{equation}

where \(m = 0,1,\ldots,N - 1\). Likewise, let
\(\widetilde{S}[ m]\) denote the N-point DFT of the noisy
signal \(\widetilde{s}[ n]\). Because the SFS outputs a
vector of estimated SNR values, \(\mathbf{FR}_{\mathrm{SNR}}\) must also be
represented as a vector for training. Therefore, the ground-truth FR SNR
of the \(k\)th tone frequency is defined as in Eq.~\eqref{eq18}.

\begin{equation} \label{eq18}
\mathbf{FR}_{\mathrm{SNR}}\left[ m_{k} \right] = \frac{\left| S\left[ m_{k} \right] \right|^{2}}{\sum_{m = 0}^{N - 1}\left| \widetilde{S}[ m] - S[ m] \right|^{2}},
\end{equation}

where \(m_{k}\) denotes the index of the frequency bin closest to
\(f_{k}\), as defined in Eq.~\eqref{eq19}.

\begin{equation} \label{eq19}
m_{k} = \mathrm{round}\left( \frac{f_{k}}{f_{s}}N \right),
\end{equation}

For effective SENet training, \(\mathbf{FR}_{\mathrm{SNR}}\) is converted to the
dB scale. The SENet loss is defined as the L1 norm between
\({\widehat{\mathbf{FR}}}_{\mathrm{SNR}}\) and \(\mathbf{FR}_{\mathrm{SNR}}\). During backpropagation, gradients are masked at frequency bins that do not
correspond to tone frequencies.

As shown in Fig.~\ref{fig3}, the SFS applies ordered-statistic constant
false-alarm rate (OS-CFAR) detection \cite{ref38} to
\(\mathbf{FR}_{\mathrm{MUL}}\) to identify tone-frequency candidates. This
process yields the TFINet tone-frequency estimates
\(\left\{ {\widehat{f}}_{{\mathrm{TFINet}}_{k}} \right\}_{k = 1}^{\widehat{K}}\)
and the estimated number of tone frequencies \(\widehat{K}\). The
SNR-based selection in Eq.~\eqref{eq20} chooses either the TFINet estimate
\({\widehat{f}}_{{\mathrm{TFINet}}_{i}}\) or the PMCD estimate
\({\widehat{f}}_{{\mathrm{PMCD}}_{i}}\). The selection criterion is the estimated
SNR \({\widehat{\lambda}}_{i}\) of the corresponding tone frequency. If
\({\widehat{\lambda}}_{i} < \lambda_{\mathrm{SNR}}\), the TFINet estimate is
selected; otherwise, the PMCD estimate is selected.

\begin{equation} \label{eq20}
\widehat{f}_{i} =
\begin{cases}
\widehat{f}_{\mathrm{TFINet},i}, & \text{if } \widehat{\lambda}_{i} < \lambda_{\mathrm{SNR}},\\
\widehat{f}_{\mathrm{PMCD},i}, & \text{otherwise.}
\end{cases}
\end{equation}

This selection is motivated by the different error characteristics of
the two estimators as a function of SNR. TFINet robustly detects
frequency bins containing tone frequencies, but its precision is limited
by the bin resolution. In contrast, PMCD provides super-resolution
frequency estimates at high SNR. At low SNR, however, the signal
subspace of PMCD is contaminated by noise, which can result in estimates
that deviate substantially from the true tone frequencies.

To determine the SNR threshold \(\lambda_{SNR}\) used by the SFS, the
breakdown SNR of PMCD is analyzed. Below a certain SNR, the signal
subspace of PMCD becomes contaminated by noise and its
frequency-estimation performance deteriorates sharply \cite{ref37}. In this study, the tone-frequency estimation error of PMCD is
compared through simulation with the Cramer-Rao lower bound (CRLB) as a
function of SNR. As the SNR increases, the PMCD error approaches the
CRLB, whereas below \(-\)6 dB the error increases sharply and departs
from the CRLB. A theoretical analysis of this behavior is provided in
the Appendix. Accordingly, the point at which the PMCD estimation error
departs from the CRLB is defined as the SNR threshold, yielding
\(\lambda_{SNR} = - 6\) dB.

\section{Experiments}

This section quantitatively evaluates the proposed SAFE through
simulations and assesses its robustness using real-world data. In
addition, an ablation study of the SFS is conducted to justify the
selected SNR threshold.

\subsection{Experimental Setup}

To train TNet in TFINet, a dataset of noisy multitone sinusoidal signals
is generated. The number of tone frequencies \(K\) ranges from 1 to 10,
and each tone-frequency component \(f_{k}\) is sampled from a uniform
distribution over \([ 0,1]\). The MLTS output
\(\mathbf{TFI}_{MLTS}\) used by TFINet is implemented using the
open-source TFI Tool \cite{ref39}. Considering that the
maximum generated tone frequency is \(f_{k} = f_{\max} = 1\), the LFM
signals are configured to be modulated up to a frequency of 1.5. The
sampling frequency \(f_{s}\) is set to 3 according to the
Nyquist-Shannon sampling theorem. This setting allows MLTS to detect
time-varying frequencies within \([ 0,1.5]\) without losing
tone-frequency information.

In the simulation, \(N = 300\), and the number of TFIs used for
summation is set to \(A = 10\). The difference between two chirp rates,
\(\Delta\gamma = \left| \gamma_{i} - \gamma_{p} \right|\) with
\((i \neq p)\), is chosen to be sufficiently large so that different
noise distributions are generated from the two chirp rates.
Specifically, LFM signals with chirp rates \(\gamma_{i}\) corresponding
to \(0^{\circ},\ \ {20/3}^{\circ},\ \ldots,\ 60^{\circ}\) are generated.
The cropping ratio \(c_{f}\) is set to \(f_{s}/2f_{\max} = 1.5\) so that
only samples corresponding to the tone-generation interval
\([ 0,1]\) are retained.
The training and test signals \(\widetilde{s}[ n]\) are
generated over the SNR range \([ - 10,\ 10]\) dB in 1-dB
increments, with 150 signals generated at each SNR. Of the complete
dataset, 80\% is used for training and 20\% for testing. Because SENet
generates \(\mathbf{FR}_{\mathrm{SNR}}\) from the TNet output
\(\mathbf{FR}_{\mathrm{MUL}}\), SENet is trained after TNet.

Model training is performed in an Ubuntu environment using an RTX 3090
GPU and PyTorch 1.13.0. TNet is trained for 100 epochs with a batch size
of 10, a learning rate of 0.001, and the AdamW optimizer \cite{ref40}. SENet for SNR estimation is trained for 50 epochs using the same optimizer and learning rate with a batch size of 64.

\subsection{Evaluation Metrics}

To quantitatively evaluate SAFE, this study follows Izacard et al.~\cite{ref8} and uses the false negative rate (FNR) and nearest-neighbor
root-mean-square error (NN-RMSE). FNR reflects the reliability of
frequency detection, whereas NN-RMSE reflects frequency-estimation
precision. The false positive rate (FPR) is additionally used to assess
false-alarm performance. Together, these three metrics provide a
comprehensive evaluation of SAFE.

FNR quantifies the proportion of ground-truth frequencies missed by a
frequency estimator. Let the ground-truth tone-frequency set be
\(F = \{ f_{1},\ldots,f_{K}\}\) and the estimated tone-frequency set be
\(\widehat{F} = \{{\widehat{f}}_{1},\ldots,{\widehat{f}}_{\widehat{K}}\}\).
An estimated frequency is regarded as a true positive (TP) if it
satisfies the condition in Eq.~\eqref{eq21} within the frequency-error tolerance
\(\epsilon\).

\begin{equation} \label{eq21}
TP = \sum_{i = 1}^{K}{\mathbf{1}\left( \min_{{\widehat{f}}_{j} \in \widehat{F}}\left| f_{i} - {\widehat{f}}_{j} \right| \leq \epsilon \right)},
\end{equation}

where \(\mathbf{1}( \cdot )\) denotes the indicator function, and the
error tolerance is set to \(\epsilon = f_{s}/2N\). The FNR \(p_{\mathrm{FN}}\)
is defined as in Eq.~\eqref{eq22}.

\begin{equation} \label{eq22}
p_{\mathrm{FN}} = 1 - TP/|F|,
\end{equation}

The number of false positives (FPs) is derived from Eq.~\eqref{eq21}, as given in
Eq.~\eqref{eq23}.

\begin{equation} \label{eq23}
FP = \sum_{{\widehat{f}}_{j} \in \widehat{F}}^{}{\mathbf{1}\left( \min_{f_{i} \in F}\left| {\widehat{f}}_{j} - f_{i} \right| > \epsilon \right)},
\end{equation}

Using Eq.~\eqref{eq23}, the FPR \(p_{\mathrm{FP}}\) is defined as in Eq.~\eqref{eq24}.

\begin{equation} \label{eq24}
p_{\mathrm{FP}} = FP/\left| \widehat{F} \right|,
\end{equation}

The NN-RMSE is defined as the root mean square of the nearest-neighbor
distances between the estimated tone-frequency set \(\widehat{F}\) and
the ground-truth tone-frequency set \(F\), as given in Eq.~\eqref{eq25}.

\begin{equation} \label{eq25}
d(\widehat{F},F) = \sqrt{\frac{1}{|\widehat{F}|}
\sum_{\widehat{f}_j \in \widehat{F}} \min_{f_i \in F}
|\widehat{f}_j - f_i|^2}.
\end{equation}

\subsection{Experimental Results}

This subsection compares the FNR and NN-RMSE of SAFE and existing
frequency-estimation methods over a wide SNR range. Performance is
evaluated both when the number of tones \(K\) is known and when \(K\) is
unknown, as is typically the case in practical environments.

\begin{figure}[t]
  \centering
  \begin{subfigure}[t]{0.48\textwidth}
    \includegraphics[width=\linewidth]{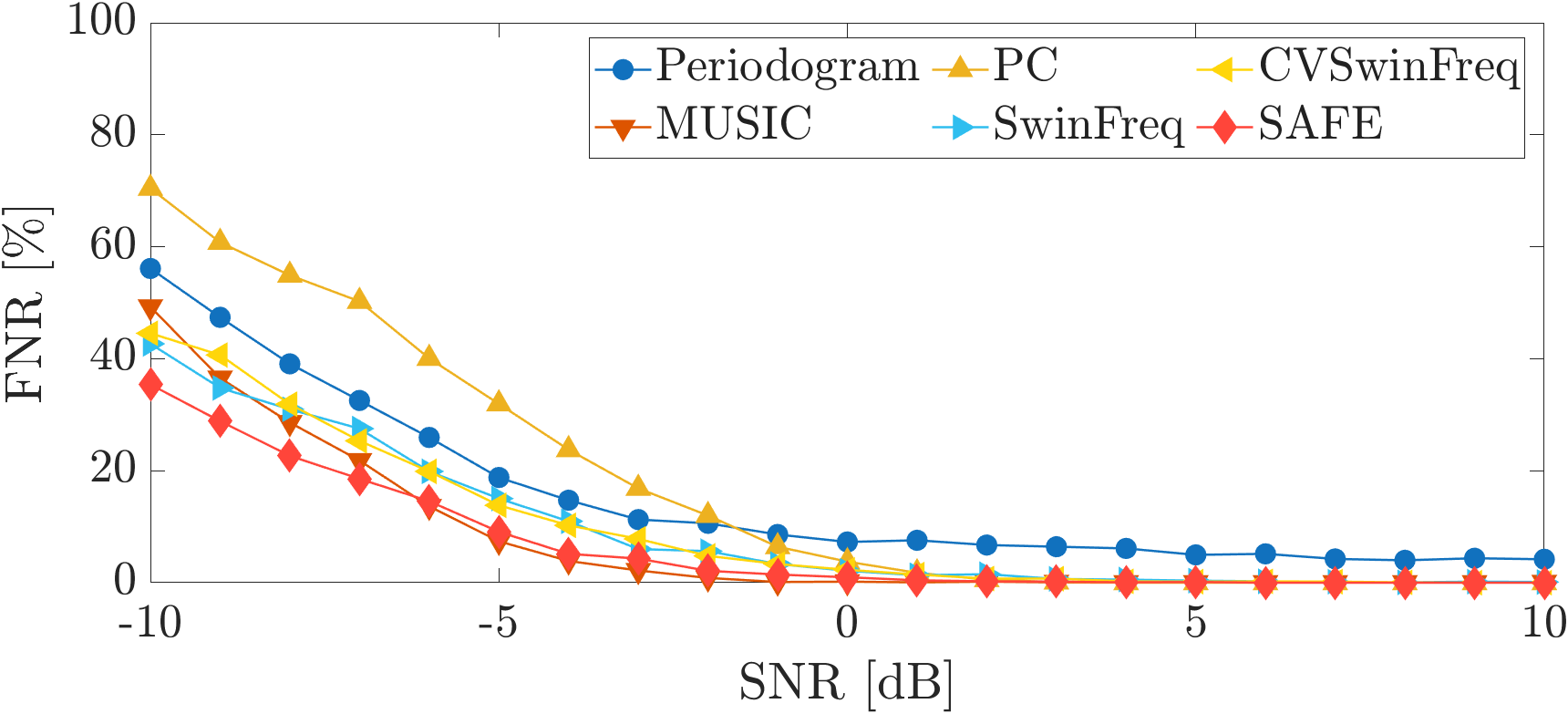}
    \caption{FNR with known $K$}
  \end{subfigure}

  \begin{subfigure}[t]{0.48\textwidth}
    \includegraphics[width=\linewidth]{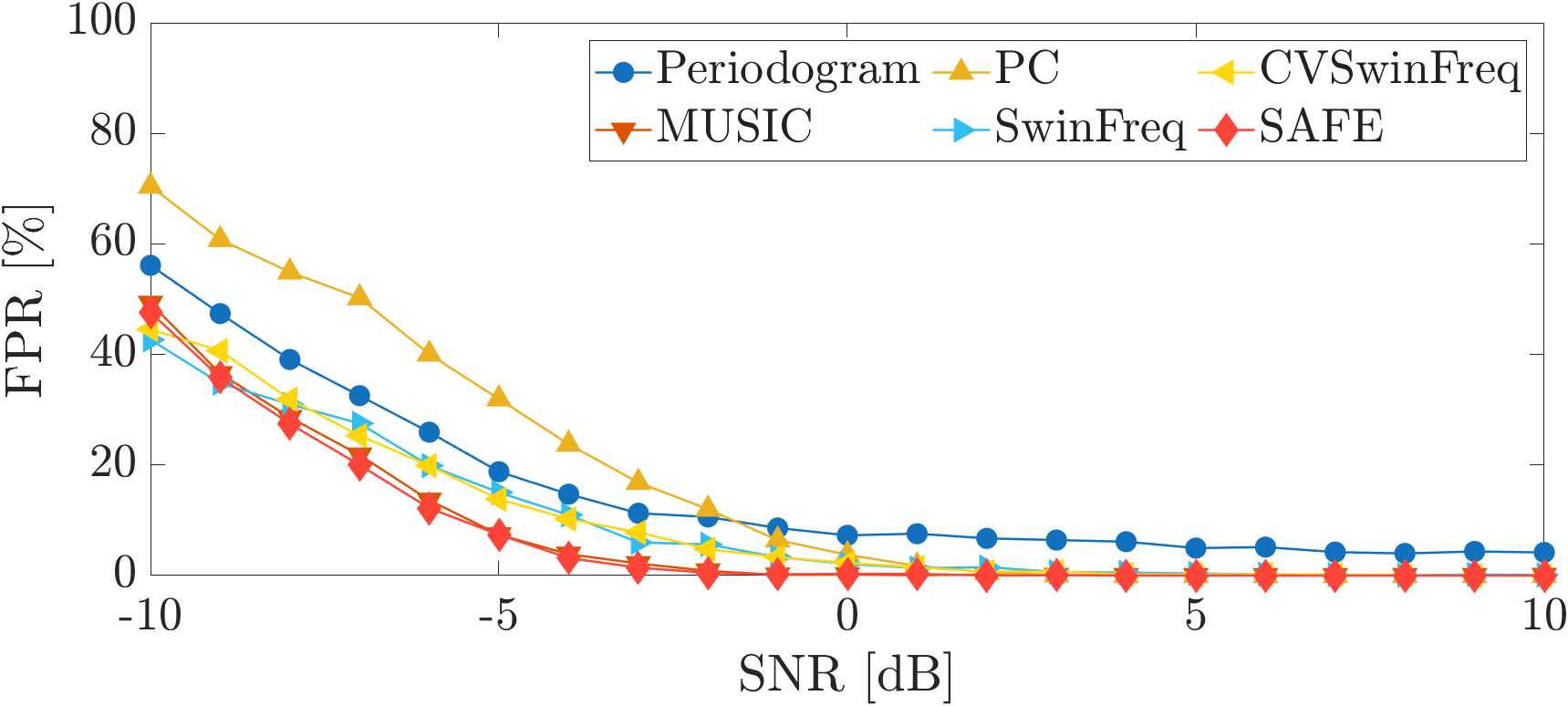}
    \caption{FPR with known $K$}
  \end{subfigure}

  \begin{subfigure}[t]{0.48\textwidth}
    \includegraphics[width=\linewidth]{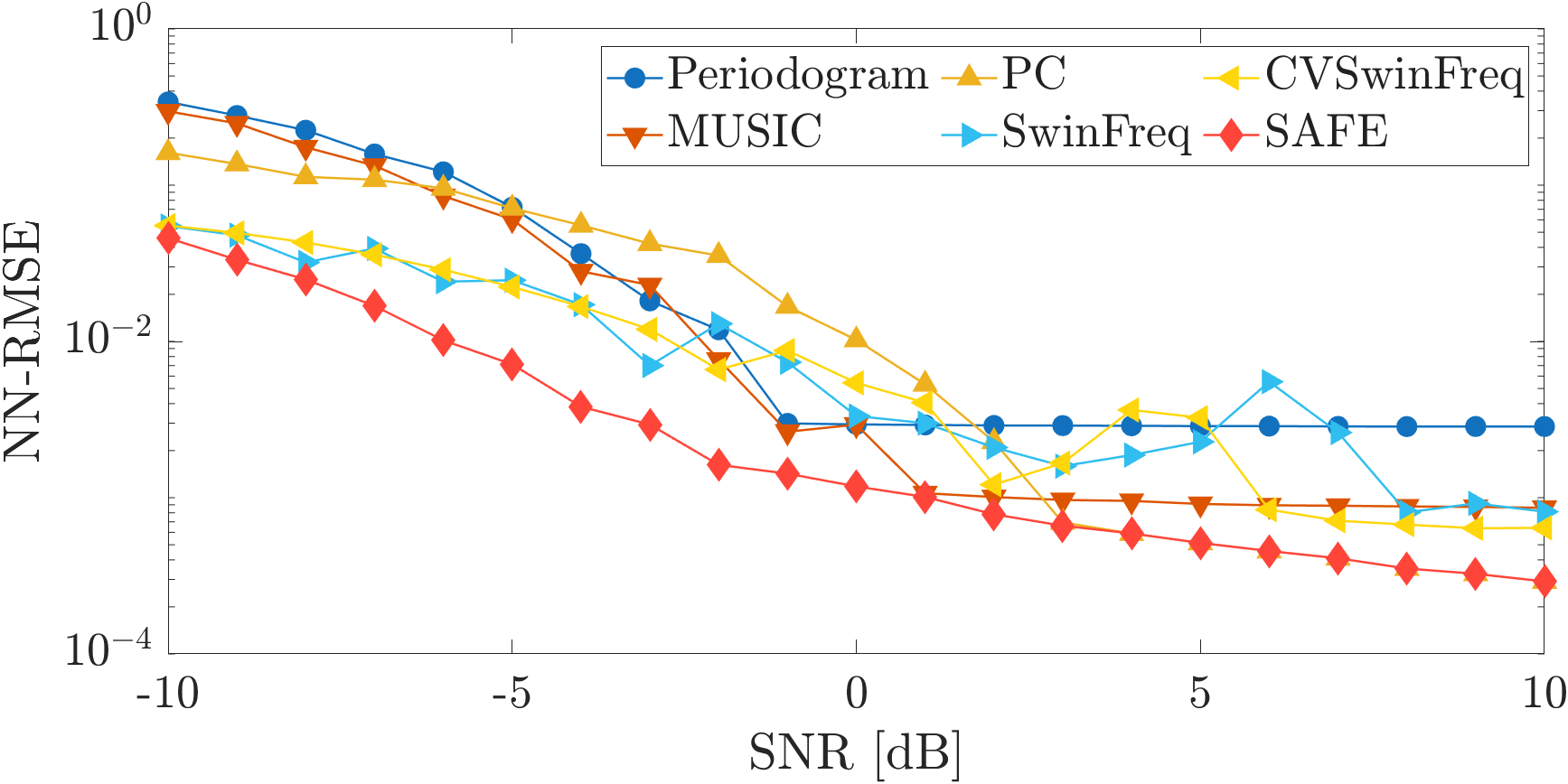}
    \caption{NN-RMSE with known $K$}
  \end{subfigure}

  \caption{
    FNR, FPR and NN-RMSE of the SAFE and existing methods for known $K$.
  }
  \label{fig4}
\end{figure}

\subsubsection{Performance With Known $K$}

First, when the number of tone frequencies \(K\) is known, SAFE is
compared with representative nonparametric, parametric, and DL-based
frequency-estimation methods. The baselines include the periodogram
\cite{ref13}, MUSIC \cite{ref15}, PMCD \cite{ref14,ref25}, SwinFreq, and CVSwinFreq \cite{ref28}.

Figs.~\ref{fig4}(a) and \ref{fig4}(c) show the FNR and NN-RMSE of the frequency-estimation
methods over a wide SNR range. In the low-SNR range (i.e.,
\(\mathrm{SNR} \leq 0\) dB), SAFE outperforms all existing methods. In
particular, MUSIC, the best-performing baseline in this range, achieves
an average FNR of 14.95\%, whereas SAFE achieves 13.00\%, corresponding
to an approximately 13.04\% reduction relative to MUSIC. In terms of
NN-RMSE, SAFE achieves a value 56.67\% lower than that of SwinFreq, the
best-performing baseline. This improvement results from the SFS
selecting TFINet, the robust frequency estimator, at low SNR. In the
high-SNR range (i.e., \(\mathrm{SNR} > 0\) dB), PMCD achieves the lowest NN-RMSE
as the SNR increases. SAFE also attains NN-RMSE values comparable to
those of PMCD because the SFS selects the super-resolution PMCD
estimator in this range. As shown in Fig.~\ref{fig4}(b), SAFE exhibits a higher
FPR than SwinFreq and CVSwinFreq at \(\mathrm{SNR} \leq - 8\) dB because some
noise peaks exceed the detection threshold during OS-CFAR processing and
are detected as tone-frequency candidates \cite{ref38}.

\begin{figure}[t]
  \centering
  \begin{subfigure}[t]{0.48\textwidth}
    \includegraphics[width=\linewidth]{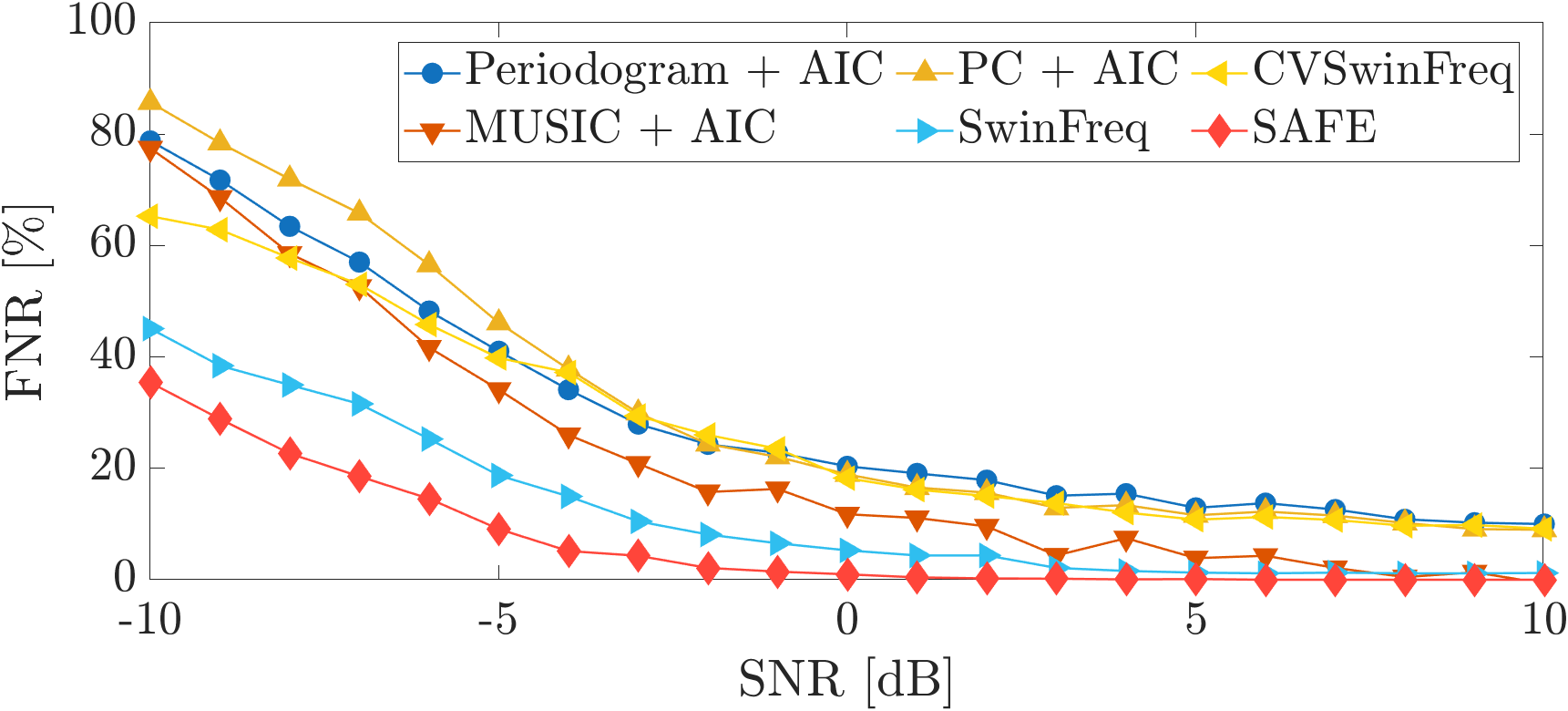}
    \caption{FNR with unknown $K$}
  \end{subfigure}

  \begin{subfigure}[t]{0.48\textwidth}
    \includegraphics[width=\linewidth]{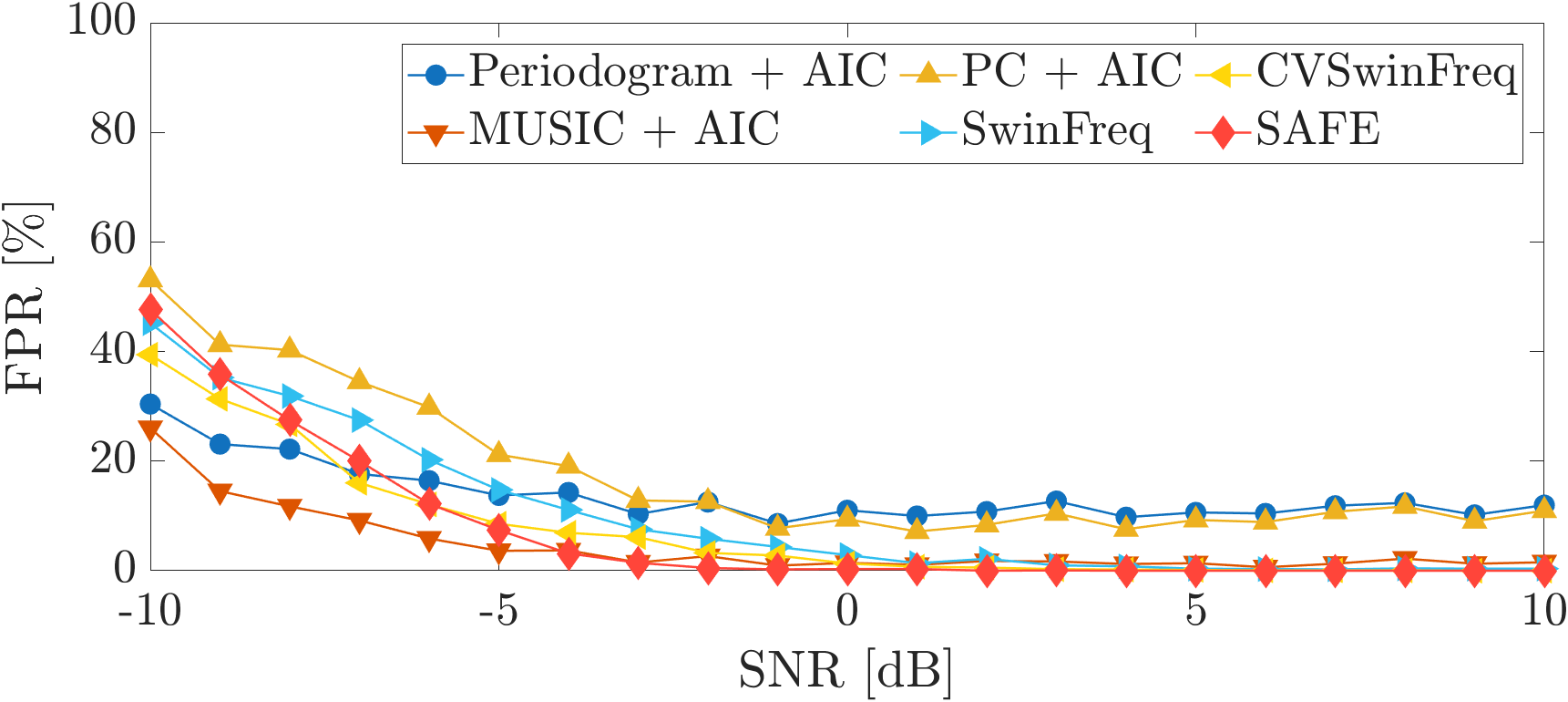}
    \caption{FPR with unknown $K$}
  \end{subfigure}

  \begin{subfigure}[t]{0.48\textwidth}
    \includegraphics[width=\linewidth]{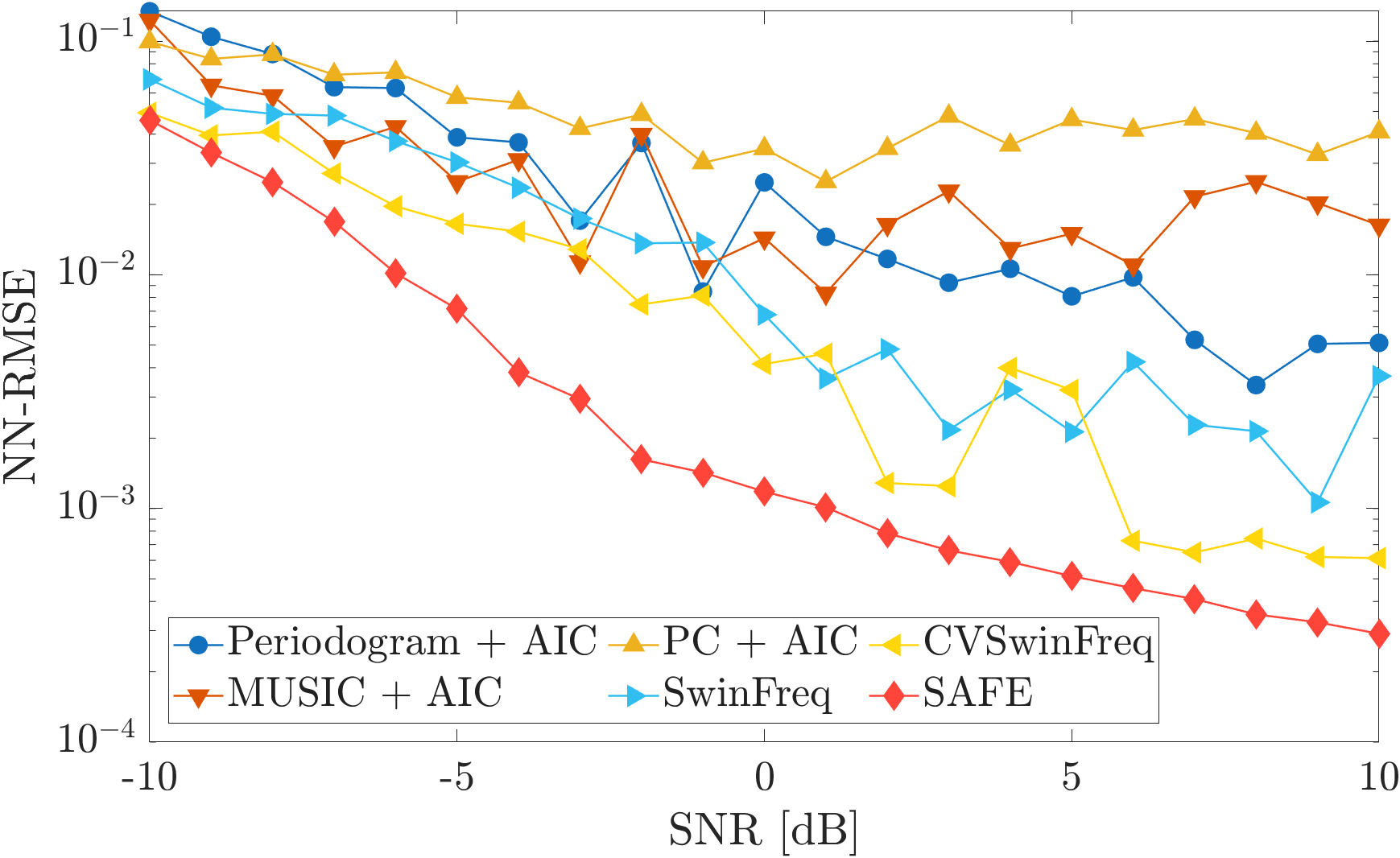}
    \caption{NN-RMSE with unknown $K$}
  \end{subfigure}

  \caption{
    FNR, FPR, and NN-RMSE of the SAFE and existing methods for unknown $K$.
  }
  \label{fig5}
\end{figure}

\subsubsection{Performance With Unknown $K$}

In most practical environments, the number of tone frequencies \(K\) is
generally unknown. For a fair comparison with SAFE, the nonparametric
and parametric methods estimate \(K\) using the Akaike information
criterion (AIC) \cite{ref41}. For the DL-based baselines, the
frequency-counting module of Izacard et al.~\cite{ref8} is combined with
SwinFreq and CVSwinFreq.

Figs.~\ref{fig5}(a) and \ref{fig5}(c) compare the FNR and NN-RMSE of the
frequency-estimation methods over a wide SNR range when the number of
tone frequencies \(K\) is unknown. In the low-SNR range, SAFE achieves
an average FNR of 13.00\%, which is 40.26\% lower than the 21.76\%
achieved by SwinFreq, the best-performing existing method. SAFE also
achieves an NN-RMSE 53.58\% lower than that of SwinFreq.

Although SAFE exhibits a higher FPR than some AIC-based existing methods
at \(\mathrm{SNR} \leq 0\) dB, it achieves the lowest FNR. This occurs because,
at low SNR, AIC-based methods may fail to distinguish weak
tone-frequency components from noise and consequently estimate a value
\(\widehat{K}\) smaller than the true \(K\). In this case, the reduced
number of detected frequencies can yield a lower FPR, but the number of
missed detections increases, resulting in a higher FNR. In contrast,
SAFE enhances weak tone-frequency components using TFINet and detects
them using OS-CFAR. As a result, some noise peaks may be included as
tone-frequency candidates and increase the FPR, but the rate of missed
true tones is reduced, yielding a lower FNR than existing methods. Thus,
by allowing a limited number of false alarms at low SNR in exchange for
higher sensitivity to weak tone frequencies, SAFE provides more robust
frequency detection when \(K\) is unknown.

\subsection{Experimental Results on Real-World Data}

This subsection evaluates the robustness of SAFE on real-world FMCW
radar data. First, a data-acquisition environment is established and
FMCW radar data are collected under multi-object detection scenarios.
The robustness of each frequency-estimation method is then evaluated on
the real-world data.

\subsubsection{Experimental Setup}

In FMCW radar, the beat frequency of the received signal is used to
determine target range \cite{ref20}. Therefore, the
multi-object ranging problem can be formulated as a multitone
frequency-estimation problem in which each tone corresponds to an
individual target. In this experiment, FMCW radar signals are acquired
from multiple human targets and the performance of the
frequency-estimation methods is compared.

The real-world data are collected in an environment containing
surrounding objects such as desks and chairs to represent a cluttered,
high-noise condition. Human targets are placed at ranges from 2 to 14 m.
To construct multitone frequency-estimation scenarios, the number of
human targets is varied as \(K = 1,\ldots,10\). For each \(K\), the
targets are randomly arranged 10 times, yielding 10 different scenarios.

Analog-to-digital converter (ADC) data are acquired using the Texas
Instruments MMWCAS-RF-EVM MIMO FMCW radar evaluation board (Texas
Instruments Inc. 2019). For each scenario described above, ADC data are
recorded for 1 s, and only a single frame at 0.5 s is used. Because the
FMCW radar consists of 12 transmitters and 16 receivers, the ADC data
are acquired from 192 virtual antennas. The resulting dataset therefore
contains \(\ 100 \times 192\  = \ 19,200\) samples, where 100 denotes
the number of target-placement cases and 192 denotes the number of
virtual antennas.

Accurate range labels are also required to train the neural networks
used by the DL-based methods. Therefore, the ranges of the randomly
placed human targets are measured using a 16-channel Velodyne 3-D LiDAR
\cite{ref43}. The measured target ranges are used as
ground truth for the tone frequencies in the frequency-estimation
problem. In addition, for each \(K = 1,\ldots,10\), the 10 generated
scenarios are randomly split into training and test sets at an 8:2
ratio.

The ADC data are preprocessed to match the input-signal conditions used
in the simulations. Because the real-world data contain 256 samples per
chirp, zero padding is applied to match the simulated input length
\((N = 300)\). An FFT is then performed along the Doppler dimension, and
the signal at the zero-Doppler index is used as the input signal to
extract the range components of the stationary human targets.

\subsubsection{Experimental Results}

\begin{figure}[t]
  \centering
  \begin{subfigure}[t]{0.48\textwidth}
    \includegraphics[width=\linewidth]{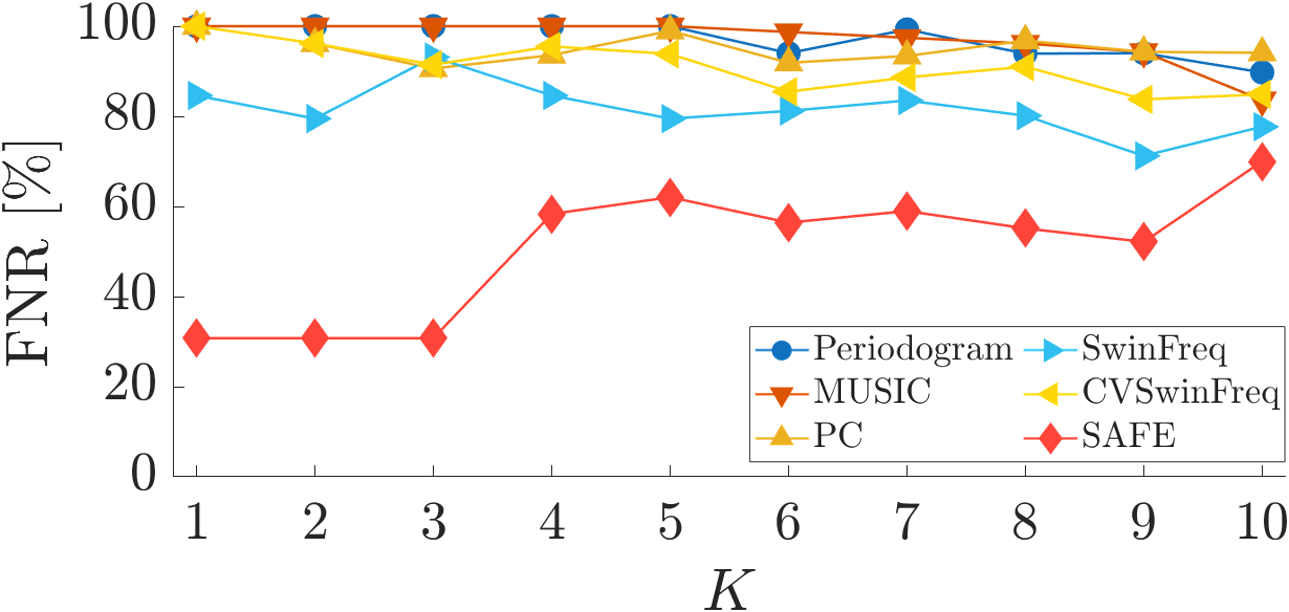}
    \caption{FNR}
  \end{subfigure}

  \begin{subfigure}[t]{0.48\textwidth}
    \includegraphics[width=\linewidth]{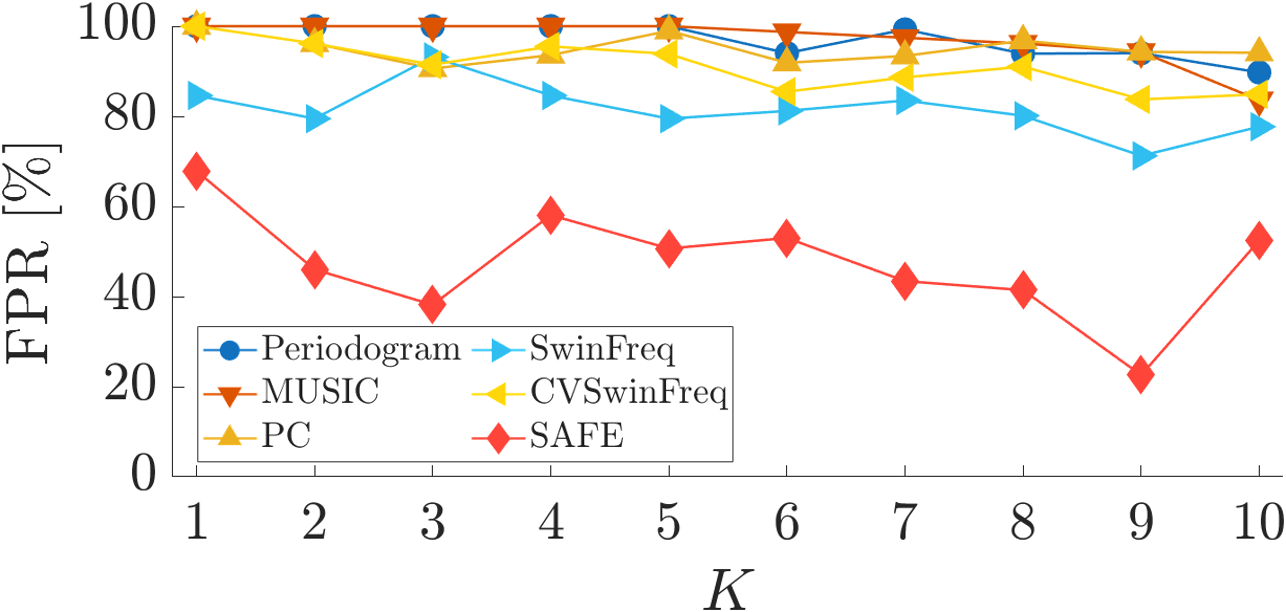}
    \caption{FPR}
  \end{subfigure}

  \begin{subfigure}[t]{0.48\textwidth}
    \includegraphics[width=\linewidth]{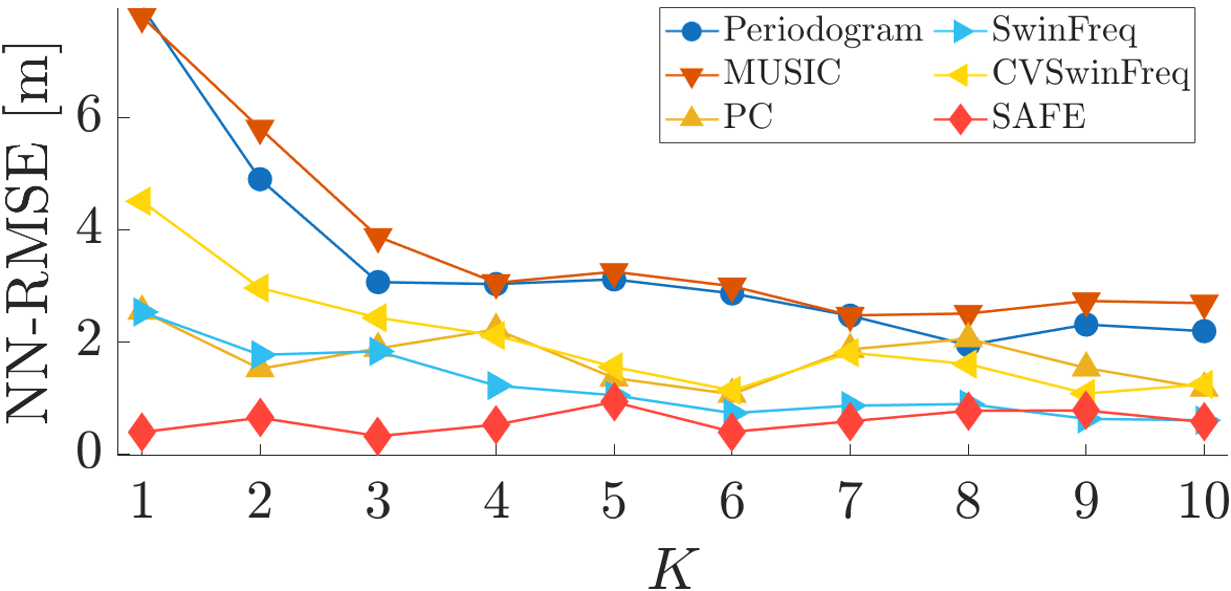}
    \caption{NN-RMSE}
  \end{subfigure}

  \caption{
    FNR, FPR, and NN-RMSE for each frequency-estimation method on the test dataset of real-world data.
  }
  \label{fig6}
\end{figure}

FNR, FPR, and NN-RMSE of the SAFE and existing methods for unknown $K$.
Figs.~\ref{fig6}(a)--\ref{fig6}(c) show the FNR, FPR, and NN-RMSE on the real-world test dataset. SAFE achieves lower FNR, FPR, and NN-RMSE than the existing
frequency-estimation methods. This result indicates that TFINet can
robustly estimate frequencies even in the presence of clutter-induced
noise components in real-world data. As shown in Fig.~\ref{fig6}(c), SAFE
achieves an average NN-RMSE of approximately 0.55 m across all numbers
of human targets.

\section{Conclusion}

This paper proposed SAFE, a hybrid frequency estimator that combines
TFINet and SFS to achieve consistent frequency estimation over a wide
SNR range. TFINet uses MLTS and TNet to enhance weak tone-frequency
components, and SAFE adaptively selects either TFINet or PMCD according
to the estimated SNR of each detected tone. This design provides robust
frequency estimation at low SNR while enabling super-resolution
frequency estimation at high SNR. Simulation results show that SAFE
outperforms the best-performing existing frequency-estimation methods in
terms of FNR and NN-RMSE even when the number of tones \(K\) is unknown.
Experiments using real-world FMCW radar data further demonstrate that
SAFE performs robust frequency estimation in both low- and high-clutter
environments, confirming its practical applicability.

\appendices
\setcounter{equation}{0}
\renewcommand{\theequation}{A\arabic{equation}}
\section{Breakdown SNR of PMCD}

This Appendix analyzes the SNR at which PMCD begins to degrade as a
function of the number of samples \(N\). Although the signal subspace
becomes increasingly contaminated by noise as the noise level increases,
PMCD estimates frequencies more robustly as \(N\) increases \cite{ref37}. The condition under which PMCD does not exhibit
performance degradation is given in Eq.~\eqref{eqA1}.

\begin{equation} \label{eqA1}
\sigma^{2} \leq \frac{d_{K}^{2}}{8(M + 1)\ln(M + 1)},
\end{equation}

where \(M = \left\lfloor (N - 1)/2 \right\rfloor\), and \(d_{K}\)
denotes the smallest singular value of the signal subspace. For
\(K = 1\), \(d_{K}\) is expressed as in Eq.~\eqref{eqA2}.

\begin{equation} \label{eqA2}
d_{K} = (M + 1)\left| a_{K} \right|.
\end{equation}

Therefore, Eq.~\eqref{eqA1} can be rewritten as Eq.~\eqref{eqA3}.

\begin{equation} \label{eqA3}
{\mathrm{SNR}}_{K} \geq \frac{8\ln(M + 1)}{M + 1},
\end{equation}

where \({\mathrm{SNR}}_{K} = \left| a_{K} \right|^{2}/\sigma^{2}\). As the number
of samples \(N\) increases, the breakdown SNR \({\mathrm{SNR}}_{K}\) of PMCD
decreases. However, because Cadzow denoising in PMCD performs SVD with
computational complexity \(\mathcal{O}\left( N^{3} \right)\), the
computational cost increases rapidly with \(N\). Therefore, considering
the tradeoff between estimation performance and computational
complexity, this study uses \(N = 300\). The theoretical breakdown SNR
is approximately \(- 5.73\) dB, which differs from
\(\lambda_{\mathrm{SNR}} = - 6\) dB by only approximately \(0.27\) dB.

\end{document}